\documentclass[aps,preprint]{revtex4}%
\usepackage{amsfonts}
\usepackage{amsmath}
\usepackage{amssymb}
\usepackage{hyperref}
\usepackage{graphicx}%
\begin{document}
\title{Circular orbits in spherically symmetric spacetimes and BSW effect with
nonzero force}
\author{H. V. Ovcharenko}
\affiliation{Department of Physics, V.N.Karazin Kharkov National University, 61022 Kharkov, Ukraine}
\affiliation{Institute of Theoretical Physics, Faculty of Mathematics and Physics, Charles
University, Prague, V Holesovickach 2, 180 00 Praha 8, Czech Republic}
\author{O. B. Zaslavskii}
\affiliation{Department of Physics and Technology, Kharkov V.N. Karazin National
University, 4 Svoboda Square, Kharkov 61022, Ukraine}

\begin{abstract}
We consider circular particle motion under the action of an unspecified force
in a static spherically symmetric spacetime. We derive the machinery that
allows one to find the force acting on a circular particle and deduce whether
its position is stable or not. This also allows one to extend the definition
of ISCO to the case of a non-zero external force. By conducting the
near-horizon expansion, we obtain that for any non-extremal black holes the
acceleration diverges, while for extremal ones it is finite. Applying the
derived machinery to the case of the Schwarzschild metric assuming that a
force is constant, we scrutiny how the number of orbits for a given force
depends on its value. In particular, if a force is big enough, an additional
branch of solutions appears that was absent in the case of geodesic motion.
Then, for various circular orbits, we numerically investigate their stability
and the position of the ISCO particles depending on the external force. In
particular, we show that the force allows ISCO particles to become closer to
the horizon, but they cannot reach it. A similar problem is solved for the
Reissner-Nordstrom (RN) metric and uncharged particles. It appears that for
the near-extremal and extremal RN black holes, there exist near-horizon circle
trajectories (in contrast to the nonextremal case), however, they are
unstable. Analysis of the ISCO particles in the RN case gives qualitatively
the same results as for Schwarzschild. In addition, high-energy particle
collisions of circular particles are considered, and it is found that they may
give an increase in the collisional energy only for near-horizon circular
orbits that exist only in near-extremal and extremal cases.

\end{abstract}

\maketitle
\tableofcontents

\newpage

\section{Introduction}

A freely moving observer is one of basic notions in both flat and curved
space-times. Meanwhile, next, a more complicated kind of an observer is an
accelerated one. Even in a simplest case of uniformly accelerated motion in
the flat space-time it leads to highly nontrivial consequences both in
geometrical properties of a frame (where it is connected with the appearance
of a horizon) and quantum physics (where it is responsible for the Unruh
effect). The situation becomes more interesting if such a motion is performed
in the curved background. It gives rise to interplay between dynamic and
geometric features. As an example, one can point out the analogue of the
Rindler horizon in the Schwarzschild space-time \cite{rs1}, \cite{rs2}.

Investigation of properties of such trajectories is of big theoretical and
methodical interest. There is a wide area where accelerated observers are
physically relevant. This includes studies of high energy particle collisions
with the generalization of the Ba\~{n}ados-Silk-West (BSW) effect \cite{ban}.
In recent works, we built an almost full theory of such an effect where
particles are not free (in contrast to \cite{ban}) but experience the action
of some unspecified force \cite{gen}. Meanwhile, our construction has a
serious gap since we concentrated on a particle falling towards a black hole
and did not consider the circle trajectories. To fill this gap, it is
necessary to study properties of circle trajectories in general setting, with
a force taken into account. To begin with, we consider here spherically
symmetric space-times such as the Schwarzschild and Reissner-Nordstr\"{o}m
ones. For free particles this is known and for the Schwarzschild metric is
even contained in textbooks (see problem 1 in Sec. 102 of \cite{LL}). However,
to the best of our knowledge, this was not done yet for a nonzero force. Small
perturbations of orbits in the Schwarzschild metric, including movement out of
the plane were considered recently in \cite{olek}, \cite{olek2}. We restrict
ourselves by motion within the plane but consider a force of an arbitrary value.

Thus our goal is two-fold. We expand consideration of properties of circle
orbits in the curved spherically symmetric background when particles are
accelerated. (In the present work, we restricted ourselves by the case when
there is no electromagnetic interaction between a particle and a black hole.)
Also, we consider how particle collision looks like for trajectories under
discussion and under which conditions the scenario of high energy collisions
is possible.

We would also like to stress that it is circle orbits for which probably exist
the best perspectives to register the BSW effect \cite{mummery25}. Such orbits
are relevant for collisions in the accretion disc \cite{gatesdisc}. Thus our
consideration is not only purely theoretical but has potential interest for
astrophysics. Here we make only the first step considering spherically
symmetric space-times but our approach will be generalized further to axially
symmetric and rotating space-times. It is worth noting that classification of
scenarios leading to high energy collisions near circle orbits was also
developed in \cite{gates25} but, in contrast to it, we include into
consideration a finite nonzero force of unspecified nature. Moreover, it is
such a force that enables us to find corresponding scenarios with high energy
particle collisions which would be otherwise impossible.

It is also worth noting that motion under the action of force is encountered
in the problems where self-force is relevant (see, for example,
\cite{detweiler08}, \cite{tores22}). Although we do not pretend for literal
application of our work to the corresponding subject, this can be though of at
lest as an additional motivation for our research.

The paper is organized as follows. In Sec. \ref{set} we list basic relations
for the metric and equations of particle motion in the spherically symmetric
static black hole when a force is present. In Sec. \ref{force} we give main
formulas for circle orbits. In Sec. \ref{hor} we point out the behavior of
acceleration for near-horizon circle orbits. In Sec. \ref{Sec_stability_def}
we discuss the conditions of stability and introduce the ISCO (innermost
stable circular orbit). In Sec. \ref{schw} we apply this formalism to circle
orbits of particles under the action of a constant force in the Schwarzschild
metric and investigate the position of ISCO particles depending on the force.
In Sec. \ref{rn} we consider this problem for the Reissner-Nordstr\"{o}m
metric. Further, we specify it for the extremal \ref{rnext} and near-extremal
cases \ref{rnne}. High energy collisions are discussed in Sec. \ref{high}. The
summary of main results is given in Sec. \ref{sum}.

\newpage

\section{General relations for circular particles}

\subsection{Particle motion under the action of an external force}

\label{set}

Let us consider generic spherically symmetric static metrics of the form
\begin{equation}
ds^{2}=-N^{2}dt^{2}+\dfrac{dr^{2}}{A}+r^{2}(d\theta^{2}+\sin^{2}\theta
d\varphi^{2})\text{,} \label{metr}%
\end{equation}
where $N=N(r),~A=A(r)$. Our convention in the choice of coordinates is
$x^{0}=t,~x^{1}=r,~x^{2}=\theta,~x^{3}=\varphi$. In the present paper we
discuss, as explicit examples, the metrics with $N^{2}=A$. However, bearing in
mind potential use in a more general context (generalized gravity theories,
non-trivial matter distributions around black holes, etc.) we consider below
formulas with different $N^{2}$ and $A$.

Let a particle with the energy $E$ and angular momentum $L$ move in the plane
$\theta=\frac{\pi}{2}$. If a particle is free, $E$ and $L$ are conserved since
the metric is independent of $t$ and an angle $\varphi$. In general, they are
not conserved if there is an action of the force that causes an acceleration
$a^{\mu}$. We assume that $a^{\phi}=0$ and the force depends on $r$ only.
Then, $L$ (but not $E$) is still conserved.

For a generic trajectory in the equatorial plane the four-velocity is given
by
\begin{equation}
u^{\mu}=(u^{t},u^{r},0,\mathcal{L}/r^{2}),~~~~\mathcal{L}=\frac{L}{m}
\label{4v}%
\end{equation}
here $m$ is the mass of a particle.

If the external force $F^{\mu}$ determined by the physics of the process of
interaction, acts on the particle, its trajectory is defined by the equation
\begin{equation}
F^{\mu}=ma^{\mu} \label{Newt_law}%
\end{equation}
where $a^{\mu}$ is the acceleration of the particle, given by
\begin{equation}
a^{\mu}=u^{\nu}\nabla_{\nu}u^{\mu}=\dfrac{du^{\mu}}{d\tau}+\Gamma_{\alpha
\beta}^{\mu}u^{\alpha}u^{\beta}, \label{accel}%
\end{equation}
$\tau$ is the proper time, $\Gamma_{\alpha\beta}^{\mu}$ being the Christoffel symbols.

To simplify calculations, instead of quantity $F^{\mu}$ we introduce a
quantity $f^{\mu}$ that has the dimension of acceleration and is defined by
\begin{equation}
f^{\mu}=F^{\mu}/m. \label{f}%
\end{equation}

To relate the trajectory of a particle to the external force, one has to
calculate its acceleration. Conducting the corresponding calculations using
(\ref{4v}) and (\ref{accel}), one obtains
\begin{equation}
a^{t}=\frac{u^{r}}{N^{2}}\frac{d(N^{2}u^{t})}{dr}\text{,} \label{at}%
\end{equation}%
\begin{equation}
a^{r}=Au^{t}\dfrac{d(N^{2}u^{t})}{dr}. \label{ar}%
\end{equation}

Here, for the calculation of the radial component of the four-acceleration,
one has to employ a normalization condition, while all other components of
acceleration appear to be zero. Using these expressions, one can easily
calculate the square of the acceleration
\begin{equation}
a^{2}=a_{\mu}a^{\mu}=\dfrac{A}{N^{2}}[\dfrac{d(N^{2}u^{t})}{dr}]^{2}%
(1+\frac{\mathcal{L}^{2}}{r^{2}})\text{,} \label{ac}%
\end{equation}

This equation can be integrated that gives us
\begin{equation}
mN^{2}u^{t}=\mathcal{E}+\beta(r)\equiv X\text{,}\label{X}%
\end{equation}
where $\mathcal{E}$ is a constant,
\begin{equation}
\beta(r)=m\int^{r}dr^{\prime}\frac{Nf(r^{\prime})}{\sqrt{A}\sqrt
{1+\frac{\mathcal{L}^{2}}{r^{2}}}}\text{.}\label{be}%
\end{equation}

Then, for the components of the four-velocity, we can write
\begin{equation}
mu^{t}=\frac{X}{N^{2}}\text{,}~~~mu^{\phi}=\frac{L}{r^{2}},~~~mu^{r}%
\equiv\sigma\dfrac{\sqrt{A}}{N}P\text{, }P=\sqrt{U_{eff}}, \label{ut}%
\end{equation}
where $\sigma=\pm1$,
\begin{equation}
U_{eff}=X^{2}-m^{2}N^{2}(1+\dfrac{\mathcal{L}^{2}}{r^{2}})\text{,} \label{U}%
\end{equation}
and the expression for the radial component of the four-velocity was obtained
from the normalization condition.

It follows from (\ref{at}), (\ref{ar}), (\ref{ut}) that the components of
acceleration are equal to%
\begin{equation}
a^{t}=\frac{\sigma}{m^{2}}\dfrac{\sqrt{A}}{N^{3}}P\frac{dX}{dr}\text{,}
\label{ats}%
\end{equation}%
\begin{equation}
a^{r}=\frac{AX}{N^{2}m^{2}}\frac{dX}{dr}. \label{ar1}%
\end{equation}

One can define the energy $E=-mu_{\mu}\xi^{\mu}$, where $\xi^{\mu}$ is the
Killing vector responsible for time translations. It follows from (\ref{X})
that $E=X$. If the force does not vanish, $E$ does not conserve. Meanwhile,
the quantity $\mathcal{E\,}$ is a constant (\ref{X}). This is in full analogy
with electrodynamics where the time component of the generalized momentum is
conserved but kinematic one is not.

According to (\ref{ut}), the forward-in-time condition $u^{t}>0$ gives us
\begin{equation}
X\geq0. \label{ft}%
\end{equation}

\subsection{Force acting on the circular particle\label{force}}

Our main goal is to concentrate on what happens for the circular orbits.
However, before doing this, we would like to separate formulas valid for any
turning point and circular orbits as such. The turning point $r=r_{t}$ where
$u^{r}=0$ is defined according to (\ref{ut}) by one relation%
\begin{equation}
U_{eff}(r)=0\text{.} \label{u0}%
\end{equation}

Meanwhile, for a circular orbit $r=r_{c}=const$, two conditions should be
fulfilled. The first condition is (\ref{u0}). But there is also the second one%
\begin{equation}
U_{eff}^{\prime}(r)=0.\label{u'0}%
\end{equation}

If only (\ref{u0}) is fulfilled (but not (\ref{u'0})), we deal with a turning
point. If both, this is a circle orbit.

It follows from (\ref{u0}) that for $r=r_{t}$ and $r=r_{c}$%
\begin{equation}
X(r)=mN(r)\sqrt{1+\dfrac{\mathcal{L}^{2}}{r^{2}}}.\label{z}%
\end{equation}
This equation can be also obtained from (\ref{ut}) if we put $u^{r}=0$ and
take into account the normalizaiton condition $u_{\mu}u^{\mu}=-1$. 

To make the condition (\ref{u'0}) meaningful, we must calculate $U_{eff}%
^{\prime}(r)$ at first. Calculating derivative of (\ref{U}) we have
\begin{equation}
\frac{U_{eff}^{\prime}(r)}{m^{2}}=2XX^{\prime}-\left(  N^{2}\right)  ^{\prime
}(1+\frac{\mathcal{L}^{2}}{r^{2}})+2\frac{\mathcal{L}^{2}}{r^{3}}%
N^{2},\label{u_pr_1}%
\end{equation}
eq. (\ref{u_pr_1}) being valid for any $r$.

By solving the condition $U_{eff}^{\prime}=0$ and using (\ref{z}), one
obtains
\begin{equation}
\Big(\dfrac{dX}{dr}\Big)_{r=r_{c}}=m\Bigg[\dfrac{d}{dr}\Big(N\sqrt
{1+\dfrac{\mathcal{L}^{2}}{r^{2}}}\Big)\Bigg]_{r=r_{c}} \label{x'}%
\end{equation}
In other words, we can calculate $(dX/dr)_{r=r_{c}}$ on a circular orbit
directly from (\ref{z}),
\begin{equation}
\lim_{r\rightarrow r_{c}}\frac{dX}{dr}=\frac{dX(r_{c})}{dr_{c}}\text{.}
\label{lim}%
\end{equation}

Eq. (\ref{lim}) uses eq. (\ref{u'0}) in an essential way. Therefore, for
$r=r_{t}\neq r_{c}$,
\begin{equation}
\lim_{r\rightarrow r_{t}}\frac{dX}{dr}\neq\frac{dX(r_{t})}{dr_{t}}\text{.}%
\end{equation}

It is worth noting that there is no analogue of eq. (\ref{lim}) for the second
and higher derivatives of $X$ since (\ref{lim}) is a consequence of eqs.
(\ref{u0}) and (\ref{u'0}) that contain no such derivatives:%
\begin{equation}
\lim_{r\rightarrow r_{c}}\frac{d^{2}X}{dr^{2}}\neq\frac{d^{2}X(r_{c})}%
{dr_{c}^{2}}\text{.}\label{no}%
\end{equation}

Instead, one can resort to eqs. (\ref{X}), (\ref{be}), so%
\begin{equation}
\lim_{r\rightarrow r_{c}}\frac{d^{2}X}{dr^{2}}=m(\frac{d}{dr}\frac{Nf}%
{\sqrt{A}\sqrt{1+\dfrac{\mathcal{L}^{2}}{r^{2}}}})_{r=r_{c}}\text{.}\label{2d}%
\end{equation}
Eq. (\ref{2d}) will be used further.

Relation (\ref{x'}) allows us to calculate the value of acceleration on the
circular orbit $a_{c}$ that can be obtained from (\ref{X}), (\ref{be})
according to which in the turning point $r=r_{t}$ or on a circular orbit
$r=r_{c}$
\begin{equation}
a^{t}=0,~~~a^{r}=\sqrt{A}a_{c},
\end{equation}
where
\begin{equation}
a_{c}=\frac{\sqrt{A}}{Nm}X^{\prime}\sqrt{1+\frac{\mathcal{L}^{2}}{r^{2}}%
}\text{.} \label{af}%
\end{equation}

Using eq. (\ref{x'}) for circular orbits, we can rewrite it in the form%
\begin{equation}
a_{c}=\frac{\sqrt{A}}{N}\Bigg(N^{\prime}\Big(1+\dfrac{\mathcal{L}^{2}}{r^{2}%
}\Big)-\frac{N\mathcal{L}^{2}}{r^{3}}\Bigg), \label{a}%
\end{equation}
where $r=r_{t}$ or $r=r_{c}$.

Now, we are ready to find $U^{\prime}(r_{t})$. By substitution of (\ref{x'})
into ((\ref{u_pr_1}) and using (\ref{a}), we obtain
\begin{equation}
\frac{U_{eff}^{\prime}(r_{t})}{m^{2}}=2f\frac{N^{2}}{\sqrt{A}}-\left(
N^{2}\right)  ^{\prime}\Big(1+\frac{\mathcal{L}^{2}}{r^{2}}\Big)+2\frac
{\mathcal{L}^{2}}{r^{3}}N^{2}=2\frac{N^{2}}{\sqrt{A}}(f-a_{c})\text{,}
\label{u_pr}%
\end{equation}
where $r=r_{t}$ in the right hand side. If, in addition, $r_{t}=r_{c}$, so
both conditions (\ref{u0}), (\ref{u'0}) are fulfilled, $U_{eff}^{\prime}%
(r_{c})=0$ as it should be by a very meaning of $r_{c}$.

We also wish to note that even though we used the coordinate frame to
calculate the acceleration, and found that only the radial component is
non-zero, this relation also remains true in the frame comoving with the
particle and the strength of the radial force in that frame is $a_{c}$, see
Appendix \ref{append_1}.

In our analysis, it is \textit{extremely important} to distinguish between the
two physical quantities. One of them is the external force $f(r)$ acting on a
particle. Another one is such a value of a force (or acceleration) $a_{c}$
that is required to keep a circular particle on a given radius which can be
calculated from kinematics. (Hereafter, we call a particle circular if it
moves along the circular orbit.) For an arbitrary trajectory, $f^{r}(r)\equiv
f(r)=a(r)$ according to (\ref{f}). However, in general $a\neq a_{c}$, where
$a_{c}$ is given by (\ref{a}), (\ref{af}). The equality is valid on the circle
trajectory \textit{only}, where the force needed to keep this particle on a
circular orbit is equal to
\begin{equation}
f(r_{c})=a_{c}.\label{fa}%
\end{equation}
This will be useful for our further calculations.

\subsection{Circle orbits near the horizon}

\label{hor}

From (\ref{a}), one can infer some conclusion concerning the properties of the
near-horizon circle trajectories. We assume that near the horizon $r_{h}$ the
Taylor expansion is valid:%
\begin{equation}
N^{2}\approx\kappa_{p}u^{p}\text{,}%
\end{equation}%
\begin{equation}
A\approx A_{q}u^{q}\text{,}%
\end{equation}
where $u=r-r_{h}$. Then, one can consider three different cases. The behavior
of $a_{c\text{ }}$near the horizon is determined by the first term in
(\ref{a}) since the second one is proportional to $\sqrt{A}$ and vanishes in
this limit. The first term is proportional to $u^{\frac{q}{2}-1}$.

\subsubsection{Non-extremal black holes}

In this case, $p=q=1$. It is seen from (\ref{a}) that acceleration diverges in
the horizon limit.

\subsubsection{Extremal black holes}

Now, $p\geq2$, $q=2.$ Then, it follows from (\ref{a}), $a_{c}$ remains finite.

\subsubsection{Ultraextremal black holes}

By definition, $p\geq2$, $q\geq3$, Then, $a_{c}\rightarrow0$.

In the particular case of an immovable particle, these properties agree with
\cite{immove}.

\subsection{Definition of stability and ISCO}

\label{Sec_stability_def}

Now, once we have found under which conditions the circular particle can be
kept at a given radius, the obvious question arises: is this position stable?
This is determined by the sign of $U_{eff}^{\prime\prime}$: if this sign is
negative, then this circular orbit is stable, while otherwise it is not. This
definition is quite natural because it follows from (\ref{ut}) that in the
first case the $u^{r}$ becomes imaginary while one tries to push the
trajectory out from the circular orbit at $r=r_{c}$, while in the second case
$u^{r}$ remains real and increases when one pushes the trajectory out of
$r=r_{c}$.

To calculate $U_{eff}^{\prime\prime}$ on a circular orbit, we differentiate
(\ref{u_pr}) with (\ref{u0}), (\ref{u'0}) taken into account and get
\begin{equation}
\frac{U^{\prime\prime}(r_{c})}{m^{2}}=Y(r_{c})\text{, } \label{uy}%
\end{equation}
where
\begin{align}
Y(r)\equiv &  2N\Big[\frac{N}{\sqrt{A}}f^{\prime}+\left(  \frac{N}{\sqrt{A}%
}\right)  ^{\prime}f+\frac{f\mathcal{L}^{2}N}{r^{3}(1+\frac{\mathcal{L}^{2}%
}{r^{2}})\sqrt{A}}\Big]+\dfrac{2f^{2}}{1+\frac{\mathcal{L}^{2}}{r^{2}}}%
\dfrac{N^{2}} {A}\nonumber\\
&  -(N^{2})^{\prime\prime}\Big(1+\frac{\mathcal{L}^{2}}{r^{2}}\Big)+4\frac
{\left(  N^{2}\right)  ^{\prime}\mathcal{L}^{2}}{r^{3}}-\frac{6N^{2}%
\mathcal{L}^{2}}{r^{4}}. \label{Y}%
\end{align}

By using (\ref{U}), (\ref{z}), (\ref{x'}) and (\ref{fa}) for $X$ and
$\frac{dX}{dr}$ and (\ref{2d}) for $\frac{d^{2}X}{dr^{2}}$, this expression
can be rewritten in the equivalent form
\begin{equation}
U_{eff}^{\prime\prime}=2m^{2}\Big[\dfrac{N^{2}}{A}(f^{\prime}-a^{\prime
})\Big]_{r=r_{c}}\text{,}\label{ufa}%
\end{equation}
where
\begin{equation}
a^{\prime}(r_{c})=\dfrac{1}{2m^{2}}\Bigg\{\dfrac{d}{dr}\Bigg(\dfrac{\sqrt{A}}{N^{2}}\dfrac{d}{dr}\Big[N^{2}\Big(1+\dfrac{\mathcal{L}^{2}}{r^{2}}\Big)\Big]\Bigg)\Bigg\}_{r=r_{c}}.\label{a'}%
\end{equation}

Eqs. (\ref{uy}) can be obtained by direct differentiation of eq. (\ref{u_pr}).
This because it is $f$ that appears in (\ref{u_pr}) and there is no assumption
$f=a_{c}$.

Thus, we see that the sign of $U_{eff}^{\prime\prime}$ depends on both
$f^{\prime}$ and $a^{\prime}$, and generally, one can make any circular
trajectory stable by adjusting the derivative of the external force. However,
if an external force is constant on a circle orbit and in its vicinity, so
$f^{\prime}(r_{c})=0$, the condition of stability is simply defined by the
sign of $a^{\prime}$ at $r_{c}$. If $a^{\prime}$ is positive there, then the
corresponding trajectory is stable, if it is negative, the trajectory is unstable.

On the border between regions of stability and instability there is an
innermost stable circular orbit (ISCO) which is characterized by the equality
$U_{eff}^{\prime\prime}=0$, so that%
\begin{equation}
f^{\prime}(r_{ISCO})=a(r_{ISCO})^{\prime}\text{,}\label{fac}%
\end{equation}
where $a(r_{ISCO})^{\prime}$ is given by (\ref{a'}). If (\ref{fac}) is
fulfilled, then instead of inequality eq. (\ref{no}) \ turns into equality.

If the external force is constant,
\begin{align}
    a(r_{ISCO})^{\prime}=0.\label{a_isco}
\end{align}
This entity is very important in the context of astrophysics.

Also we wish to notice that the expression (\ref{ufa}) can be obtained by
taking a second derivative of (\ref{U}). To obtain exactly (\ref{ufa}) one has
to be careful of the interpretation of what is considered as an external force
and what is the acceleration of the particle. Expression (\ref{ufa}) will be
obtained if one interprets the derivatives of $N^{2}(1+\mathcal{L}^{2}/r^{2})$
as the terms, related to acceleration, while the terms with the derivatives of
$X^{2}$ as the ones, related to the external force. There is nothing strange
with such an interpretation, as $X^{2}$ is related to the kinematics of the
particle itself and the corresponding derivatives represent the acceleration
of such particle (that is related to the external force by (\ref{Newt_law})),
while the derivatives of $N^{2}(1+\mathcal{L}^{2}/r^{2})$ represent the
acceleration, required to keep circular particle on a given radius. This fact
may be confirmed by calculation of the acceleration of the circular particle,
conducted in the Appendix \ref{append_1}.

\section{ Schwarzschild metric\label{schw}}

Now, let us move to the application of the general analysis of circular
trajectories under the action of the external force we have developed so far
to some specific cases. The simplest space-time is the Schwarzschild black
hole. For this space-time $A=N^{2}=1-\frac{r_{s}}{r}$. Then we obtain from
(\ref{z}) that for a circular orbit
\begin{equation}
X(r_{c})=m\sqrt{1-\frac{r_{s}}{r_{c}}}\sqrt{1+\frac{\mathcal{L}^{2}}{r_{c}%
^{2}}}.
\end{equation}

The force at $r_{c}$ is given by $f(r_{c})=a_{c}$, where according to
(\ref{a}) and employing (\ref{x'}), one obtains on a circle orbit
\begin{equation}
a_{c}=\dfrac{X}{m^{2}}\frac{\sqrt{A}}{N^{2}}\partial_{r}X=\frac{1}%
{\sqrt{1-\frac{r_{s}}{r_{c}}}}\frac{\mathcal{L}^{2}(\frac{3r_{s}}{2}%
-r_{c})+\frac{r_{s}}{2}r_{c}^{2}}{r_{c}^{4}}. \label{rad_eq_with_F}%
\end{equation}

For our future computations, it will be useful to introduce new dimensionless
quantities%
\begin{equation}
b=\dfrac{m\mathcal{L}}{r_{s}},\text{ \ \ }y=\frac{r_{c}}{r_{s}},\text{ }\alpha=a(y_{c})r_{s}.
\end{equation}

Then the dimensionless acceleration $\alpha=a_{c} r_{s}$ at a given radius
\begin{equation}
\alpha=a_{c}(y_{c})r_{s}=\sqrt{\frac{y}{y-1}}\frac{b^{2}(3-2y)+y^{2}}{2y^{4}%
}=f(y_{c})r_{s}. \label{f_y_eq}%
\end{equation}

Unfortunately, it cannot be solved analytically to find $y$. But we can carry
out some analysis of it. Hereafter, we will assume that $b\geq0$ without a
loss of generality$.$

If the force $f$ is absent, the circular orbits are defined by the condition
$b^{2}(3-2y_{c})+y_{c}^{2}=0$ that gives for geodesic circular orbits equation%
\begin{equation}
y_{c}=b^{2}\left(  1\pm\sqrt{1-\frac{3}{b^{2}}}\right)  , \label{r0co}%
\end{equation}
which is known result for the Schwarzschild space-time. The innermost stable
circle orbit (ISCO) is where both these roots are equal, namely if $b^{2}=3$.
In this case $y_{c}=3.$

In what follows, we assume that $f$ does not depend on $r$, so $\alpha$ in eq.
(\ref{f_y_eq}) is a constant. Now let us find how this situation changes with
the presence of a small force. In this case, we assume that the trajectory
radius changes slightly, namely that $y=y_{c}^{(0)}+\alpha y_{c}%
^{(1)}+o(\alpha)$ where $y_{c}^{(0)}$ is given by (\ref{r0co})$.$ Substituting
this to (\ref{rad_eq_with_F}) and expanding by small $f$, one obtains%
\begin{equation}
y_{c}^{(1)}=-\frac{4(y_{c}^{(0)})^{9/2}\left(  y_{c}^{(0)}-1\right)  ^{3/2}%
}{(y_{c}^{(0)})^{2}\left(  4y_{c}^{(0)}-3\right)  -b^{2}\left[  12(y_{c}%
^{(0)})^{2}-34y_{c}^{(0)}+21\right]  }. \label{y_co_eq}%
\end{equation}

For any finite $b$ and $f\rightarrow0$, there is one more root%
\begin{equation}
y_{c}\approx\frac{1}{\sqrt{2\alpha}} \label{small}%
\end{equation}
independently of $b$.

Eq. (\ref{y_co_eq}) fails if, without a force, a particle moves on ISCO where
$b=\sqrt{3},~y_{c}^{(0)}=3$ (in this case, the denominator in (\ref{y_co_eq})
vanishes). This means that in this case our assumption that the force and the
deviation of the trajectory are linearly dependent, ceases to be valid.
Instead, $y_{c}=3+\sqrt{\alpha}y_{c}^{(1)}+o(\sqrt{\alpha}).$ Substituting
this and $b=\sqrt{3}$ into (\ref{f_y_eq}), one obtains
\begin{align}
y_{c}^{(1)}=\pm3\cdot6^{3/4}.
\end{align}

\begin{figure}[ptb]
\centering
\includegraphics[width=1\linewidth]{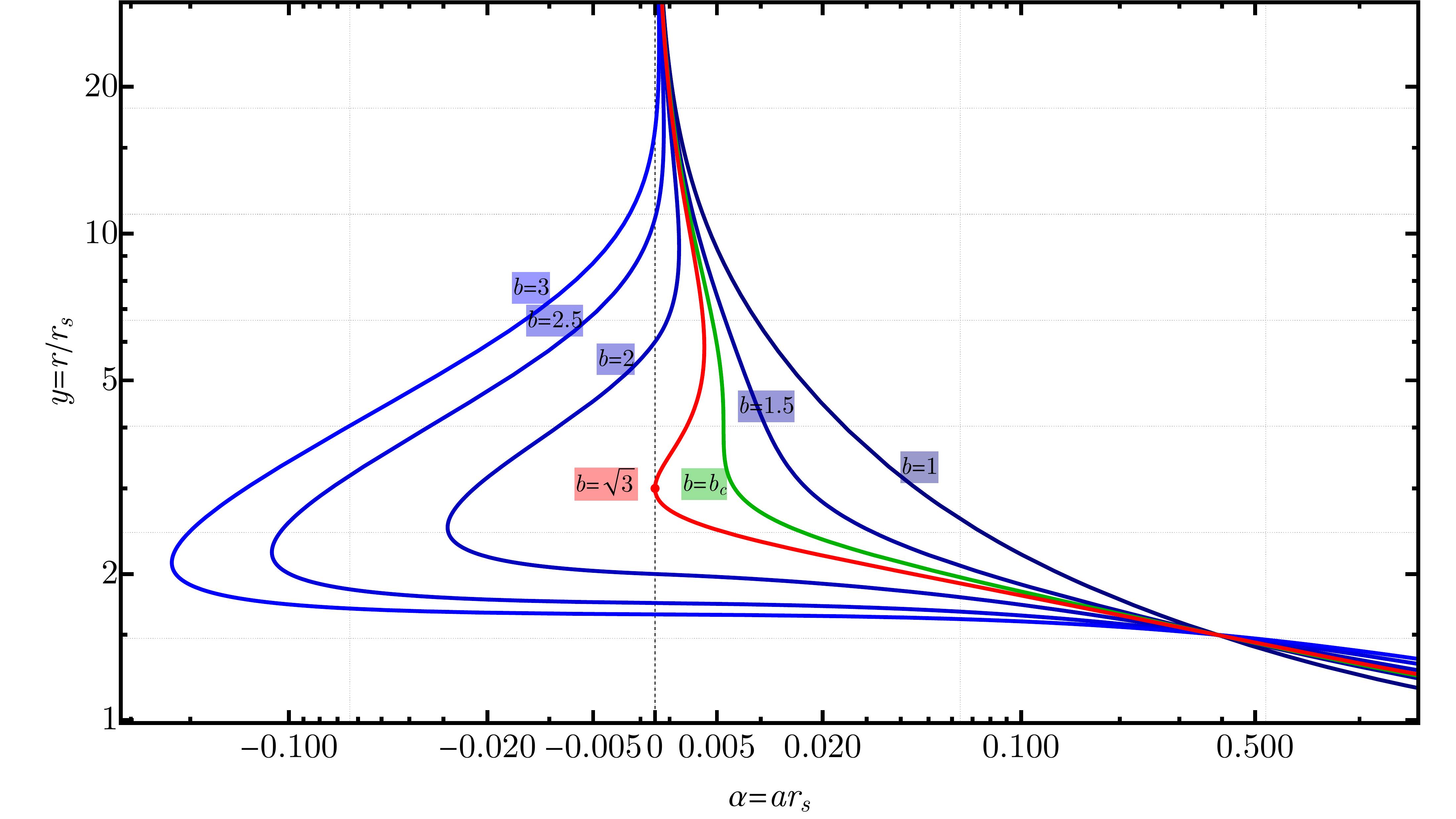} \caption{Plot showing
dependence of possible positions of circular orbits on the dimensionless
acceleration $\alpha=ar_{s}$, experienced by the circular particle at a given
dimensionless radius $y=r_{c}/r_{s}$ for various angular momenta. The red
curve represents a value of $b=\sqrt{3}$, corresponding to ISCO in the absence
of force. The plot is presented in the double logarithmic scale. Here
$b=L/r_{s}$}%
\label{fig_roots}%
\end{figure}

Notice the $\pm$ sign in this expression. It appears because the circular
orbit with $b=\sqrt{3}$ (the red curve in FIG. \ref{fig_roots}) has two close
roots for a small force. This means that for a small positive force there are
two branches in the vicinity of the ISCO.

Eq. (\ref{f_y_eq}) cannot be solved exactly but can be analyzed. First of all,
we note that for $y\rightarrow1,$ according to (\ref{f_y_eq}), $\alpha
\rightarrow\infty.$ If $y\rightarrow\infty,$ $\alpha\rightarrow0.$ Thus, we
see that by approaching the horizon from $y\rightarrow\infty$ to $y=1,$ the
force $f(y)$ required to keep a particle on a circle orbit at a given radius
increases from $0$ to $\infty.$ However, this growth may be non-monotonic. It,
obviously, affects the number of roots.

To find how many roots are present for a given force, let us investigate the
1-st derivative of the force with respect to $y$:%
\begin{equation}
\alpha^{\prime}=\left(  \frac{y}{y-1}\right)  ^{3/2}\frac{(3-4y)y^{2}
+b^{2}(21-34y+12y^{2})}{4y^{6}}. \label{f_prime}%
\end{equation}

The quantity in the numerator has real roots only if
\begin{equation}
5328b^{6}-9400b^{4}-13275b^{2}-567\geq0.
\end{equation}

This happens only if $b\geq b_{c}\approx1.64354$ (see the green curve on Fig.
\ref{fig_roots}). This means that if $b\in\lbrack0,b_{c}],$ then
(\ref{f_prime}) does not have roots and the function $\alpha(y)$ monotonously
changes from $\infty$ to $0$ (when $y$ growth from $1$ to $\infty$). Thus, if
$b\in\lbrack0,b_{c}],$ and $\alpha>0,$ there is always one root of
(\ref{f_prime}), while if $\alpha\leq0,$ there are no finite roots of
(\ref{f_prime}). For $b\in(b_{c},\infty)$ equation (\ref{f_prime}) starts to
have 2 roots (there are only two of them because the third one corresponds to
$y<1$, namely it is below the horizon and thus is not interesting for us).
This means that for $b\in(b_{c},\infty)$ there is always one local minimum and
local maximum. However, there appears to be a question about where these
minimum and maximum are located. An exact analytical answer is impossible, but
we can still do some analysis. First of all, we note that the local minimum
mentioned above becomes negative for $b\geq\sqrt{3}$ (this can be checked by
direct substitution into both \eqref{f_y_eq} and \eqref{f_prime}, see also the
red curve on Fig. \ref{fig_roots}). This means that for $b\in(b_{c},\sqrt{3})$
both local minimum and maximum are positive, while for $b\in\lbrack\sqrt
{3},\infty)$ the local minimum becomes negative. Thus the whole picture of
possible roots of (\ref{f_y_eq}) is presented in Table \ref{tab_roots}.
\begin{table}[ptb]
\centering
\begin{tabular}
[c]{|c|c|c|}\hline
$|b|\in$ & $\alpha>0$ & $\alpha\leq0$\\\hline\hline
$[0, b_{c}]$ & 1 root & No roots\\\hline
$( b_{c},\sqrt{3})$ & 1, 2 or 3 roots & No roots\\\hline
$[\sqrt{3},\infty)$ & 1, 2 or 3 roots & 1, 2 or no roots\\\hline
\end{tabular}
\caption{Number of roots of eq. (\ref{f_y_eq}) representing the possible
amount of circular orbits for a given force, depending on the dimensionless
angular momentum $b=\dfrac{L}{r_{s}}$ and force $f=\alpha/r_{s}$. For positive
$\alpha$, two roots may be achieved if one of them is the local maximum, for
negative $\alpha$'s, 1 root is the local minimum.}%
\label{tab_roots}%
\end{table}

These analytical results are supported by numerical analysis presented on Fig.
\ref{fig_roots}. Results on this plot are in full correspondence with our
analytical analysis: for angular momenta smaller than $b_{c}$, the
corresponding function $\alpha(y)$ is monotonous. Above this value it is not
monotonous and has only one minimum and one maximum. For $b=\sqrt{3}$ there
appears a first circular orbit with $f=0$ (corresponding to ISCO for free
particles), and for $b>\sqrt{3}$ there are two trajectories that can be
achieved for $f=0$.

\begin{figure}[ptb]
\centering
\includegraphics[width=1\linewidth]{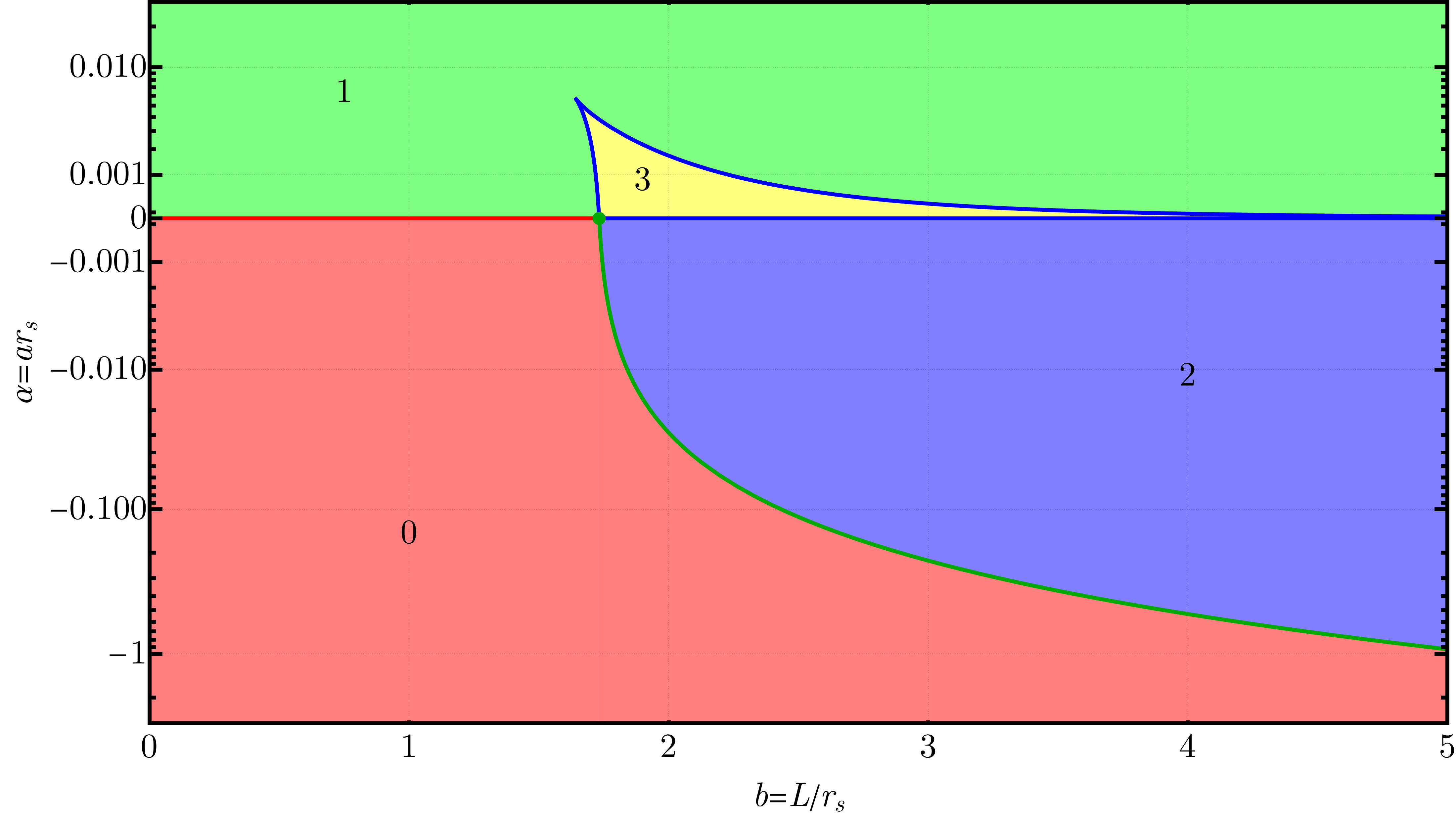} \caption{Plot showing the
number of circular orbits for various accelerations and angular momenta. Here,
red color means that at a given point, there are no roots, green--there is 1
root, blue--there are 2 roots, yellow--there are 3 roots. The green dot where
all regions meet corresponds to ISCO.}%
\label{fig_2}%
\end{figure}

In addition, we show Fig. \ref{fig_2}, where we plot the number of circular
orbits for various values of force and angular momentum calculated
numerically. Our analysis correlates with this diagram.

\begin{figure}[ptb]
\centering
\includegraphics[width=0.8\linewidth]{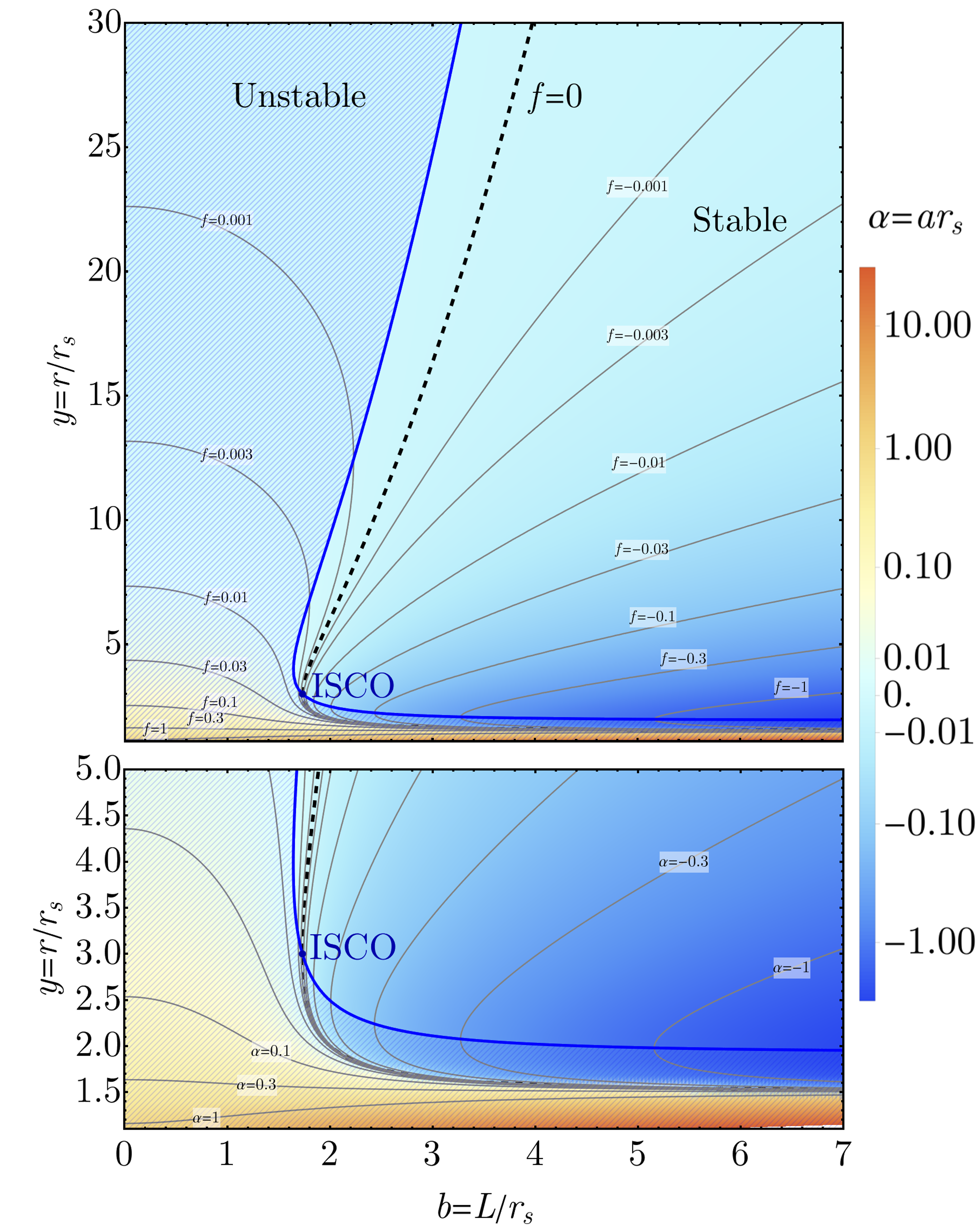} \caption{Diagram showing the
acceleration $\alpha=ar_{s}$ experienced by a particle for various radial
positions and angular momenta. The corresponding color represents the force
acting on such a particle that can be found from the legend on the right. The
dashed curve represents circular orbits that can be obtained without any
force. The hatched region represents an unstable region, the blue curve--the
boundary of the unstable region. Blue point, that is, the cross-section of the
dashed and blue curves, represents the ISCO without a force. The panel below
shows the smaller part of the same diagram.}%
\label{fig_3}%
\end{figure}

Also, we would like to investigate the stability of circular orbits for such
observers. We say that the circular orbit is stable if the particle does not
fall into the black hole under small perturbations of the radius of the
circular orbit. The conditions required for this to happen were discussed in
detail in Sec. \ref{Sec_stability_def}. Hereafter, \textit{we assume that the
external force }$f$ \textit{remains the same under perturbations, thus}
$f^{\prime}=0$. Then, the sign of $U_{eff}^{\prime\prime}$ given by
(\ref{ufa}), is defined solely by the sign of the derivative of the
acceleration $a^{\prime}$. If it is positive, then the orbit is stable,
otherwise it is not stable. The corresponding stability diagram is presented
in Fig. \ref{fig_3}, where the unstable region is hatched. Notice that the
intersection of the edge of the region of stability (blue curve) and the curve
of zero force is exactly the ISCO without the force. This is in correspondence
with the usual properties of ISCO being the closest to the horizon stable
circular orbit.

\begin{figure}[ptb]
\centering
\includegraphics[width=1\linewidth]{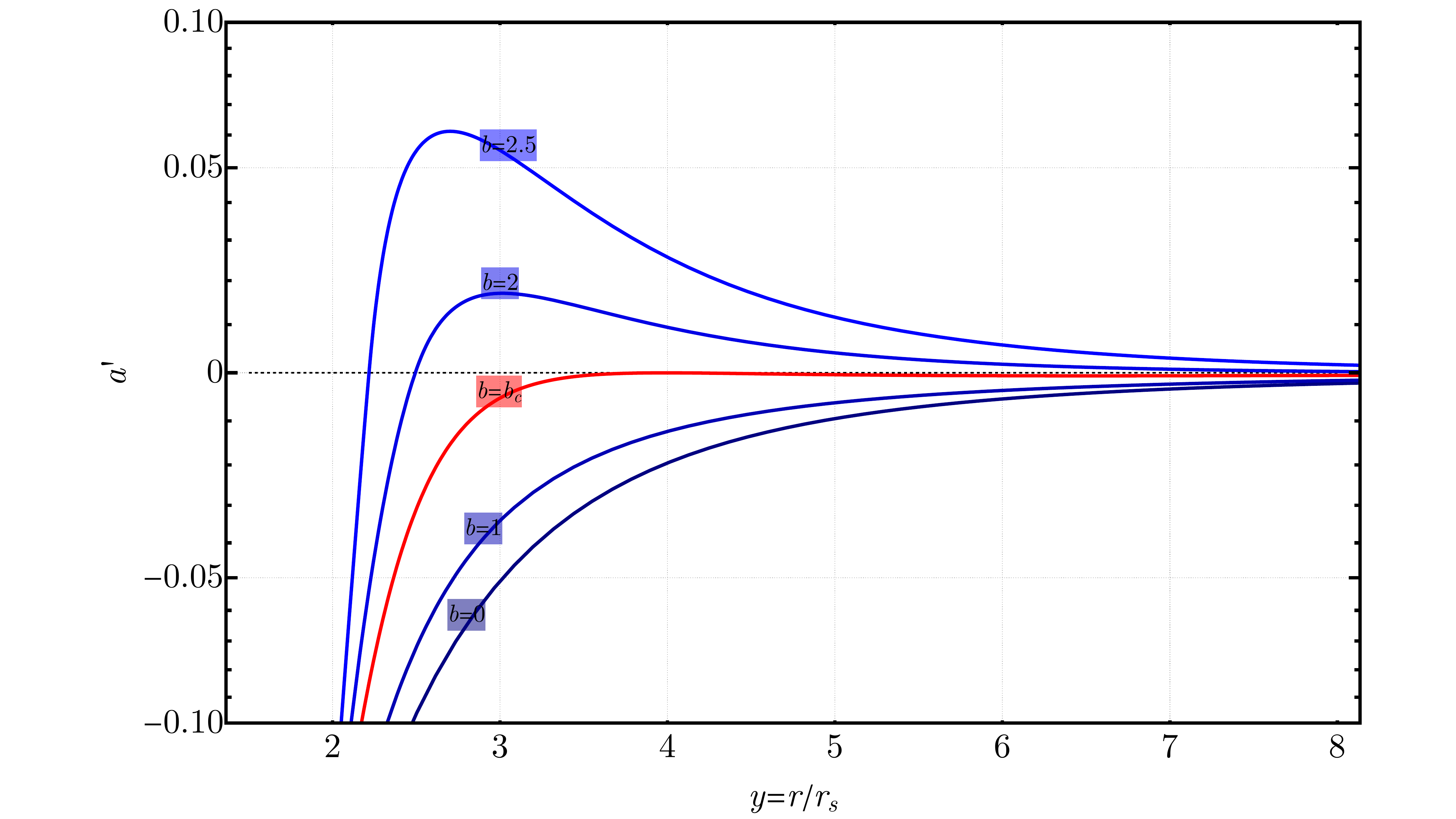} \caption{Dependence of
$a^{\prime}=da/dr$ on $y=r/r_{s}$ for various angular momenta $b=L/r_{s}$.}%
\label{fig_4}%
\end{figure}

However, this diagram may not be informative, because of this, we plot the
dependence of $a^{\prime}$ on the position of the circular orbit for various
angular momenta, see Fig. \ref{fig_4}. Indeed, one may see, that for $b<b_{c}$
equation $a^{\prime}=0$ does not have any roots, while for $b\geq b_{c}$ it
has 2 roots that correspond to the local minimum and the local maximum. As can
be seen from both \ref{fig_3} and \ref{fig_4}, any circular orbit close to the
horizon is unstable.

In \cite{F} it was shown that the analogue of the BSW effect is possible for
the Schwarzschild metric due to the presence of a force. However, this
requires that particle approach the horizon. It does not apply to circle
orbits since, as we saw in Sec. \ref{hor}, circle orbits do not exist near the
nonextremal horizon for a finite force.

At the end, one may be interested in the position of the ISCO particles
depending on the external force. The physical relevance of such an
investigation is quite clear, as the ISCO trajectories in some models
represent the edge of the accretion disc, and the information on how the force
changes the ISCO may be even used for direct observations.

\begin{figure}[ptb]
\centering
\includegraphics[width=1\linewidth]{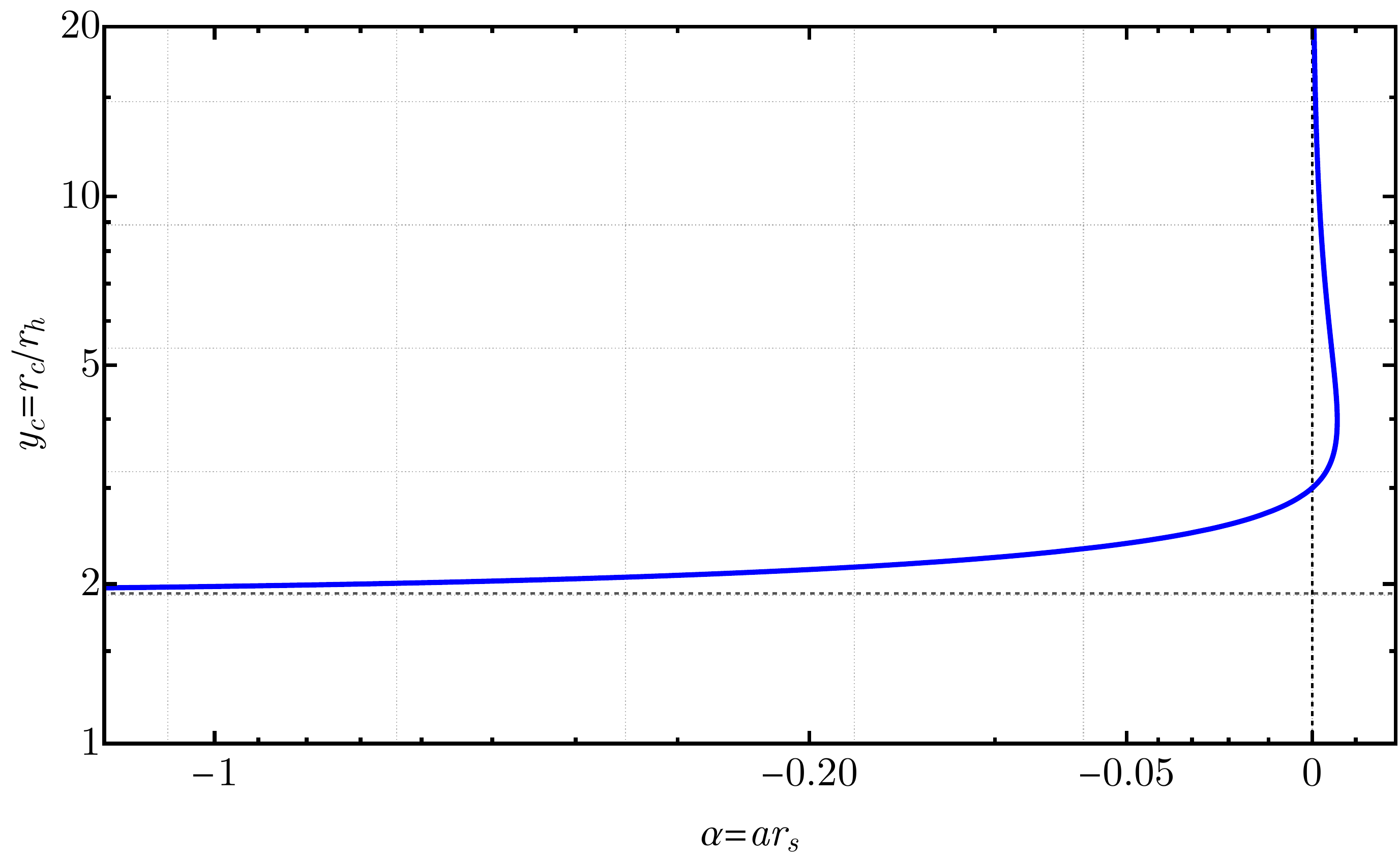}
\caption{Position of ISCO depending on the force acting on the particle in
Schwarzschild spacetime.}%
\label{pic1_isco}%
\end{figure}

To obtain the position of ISCO depending on the external force, we proceed in
such a way. For this, we employ the condition that for the ISCO particles
$a^{\prime}=0$ (see (\ref{a_isco})). From this condition, we, using
(\ref{f_y_eq}), can explicitly find the angular momentum $b$ for which the
condition $a^{\prime}=0$ is satisfied. Then, by substituting this value of $b$
into (\ref{f_y_eq}), we obtain how the force for ISCO depends on its position
$y$. Knowing this dependence, we can \textquotedblleft
invert\textquotedblright\ it to obtain the dependence of the position of ISCO
on force. The result of the application of the aforementioned algorithm is
presented in Fig. \ref{pic1_isco}. In this picture, there are several
interesting properties. First of all, we notice that in the case of zero force
(vertical dashed line), the ISCO is placed at $y=3$ as it has to be. When the
force is decreasing (which means that we are \textquotedblleft
pushing\textquotedblright\ a particle towards a black hole), the position of
ISCO becomes closer to the horizon; however, it never reaches the horizon. The
limiting position of the ISCO is $y=\dfrac{17+\sqrt{37}}{12}$ (the horizontal
dashed line in Fig. \ref{pic1_isco}), the proof of this fact can be found in
Appendix \ref{app}. This interesting property can be explained in such a way:
when the external force pushes a particle towards the horizon, the particle
has to increase its angular momentum to balance both the gravitational and the
external force. However, after some limit, this \textquotedblleft
balancing\textquotedblright\ cannot be done and the particle falls onto a BH.
This situation is somehow analogous to the case of usual ISCO without the
force, when, after some limit $y=3$ the centrifugal force cannot balance the
gravitational one, but the non-zero force allows to \textquotedblleft
widen\textquotedblright\ this range and push the particle closer.

However, when one increases the force (which means ``pulling'' a particle out
of a black hole), some interesting phenomena occur. Namely, it appears that
for a given force there appear \textit{two} radii at which an ISCO particle
may exist. If one increases the force even more, the ISCO trajectories cannot
exist at all (this limiting value of the force is $\alpha=0.00562435$, and it
corresponds to $y=3.97897524$). There also exists an explanation for this
phenomenon, namely, if we are ``pulling'' a particle, its radial position
increases, and to remain an ISCO, a particle has to have lower angular
momentum. After some limiting value of the force, the angular momentum cannot
drop lower, and thus the ISCO trajectories cannot exist for a larger force.

\newpage

\section{Reissner-Nordstr\"{o}m metric}

\subsection{Nonextremal case}

\label{rn}

The next case we would like to investigate is the motion of uncharged
particles around the Reissner-Nordstr\"{o}m black hole. For this space-time
the metric functions in (\ref{metr}) are given by
\begin{equation}
A=N^{2}=\dfrac{(r-r_{+})(r-r_{-})}{r^{2}}\text{,} \label{Arn}%
\end{equation}
where $r_{\pm}=M\pm\sqrt{M^{2}-Q^{2}}$, $M$ being the mass of a black hole,
$Q$ its charge.

Description of general properties of neutral particles trajectories in this
metric can be found in \cite{pugliesen}. Motion of charged ones which
experience Coulomb interaction with a black hole is considered in
\cite{pugliese17}, \cite{Schroven2021}. Meanwhile, we investigate circle
trajectories of neutral particles but moving under the action of an
unspecified nature. Moreover, we make the main emphasis on the action of a
constant force, so in contrast to the Coulomb one, it does not depend on a coordinate.

Substituting these expressions to (\ref{af}) and introducing new parameters
$\alpha=a_{c}r_{+},~b=L/r_{+},~y=r/r_{+},~y_{-}=r_{-}/r_{+}$, one obtains
\begin{equation}
\alpha=\dfrac{y^{3}(1+y_{-})-2y^{2}(y_{-}+b^{2})+3b^{2}(1+y_{-})y-4y_{-}b^{2}%
}{2y^{4}\sqrt{(y-1)(y-y_{-})}} \label{f_y_rn}%
\end{equation}

One can notice that in the limit $y_{-}\rightarrow0$ one directly recovers
(\ref{f_y_eq}). However, generally, the additional parameter $y_{-}$ makes the
analytical analysis extremely complicated. Because of this, we will perform
numerical analysis only.

\begin{figure}[ptb]
\centering
\includegraphics[width=1\linewidth]{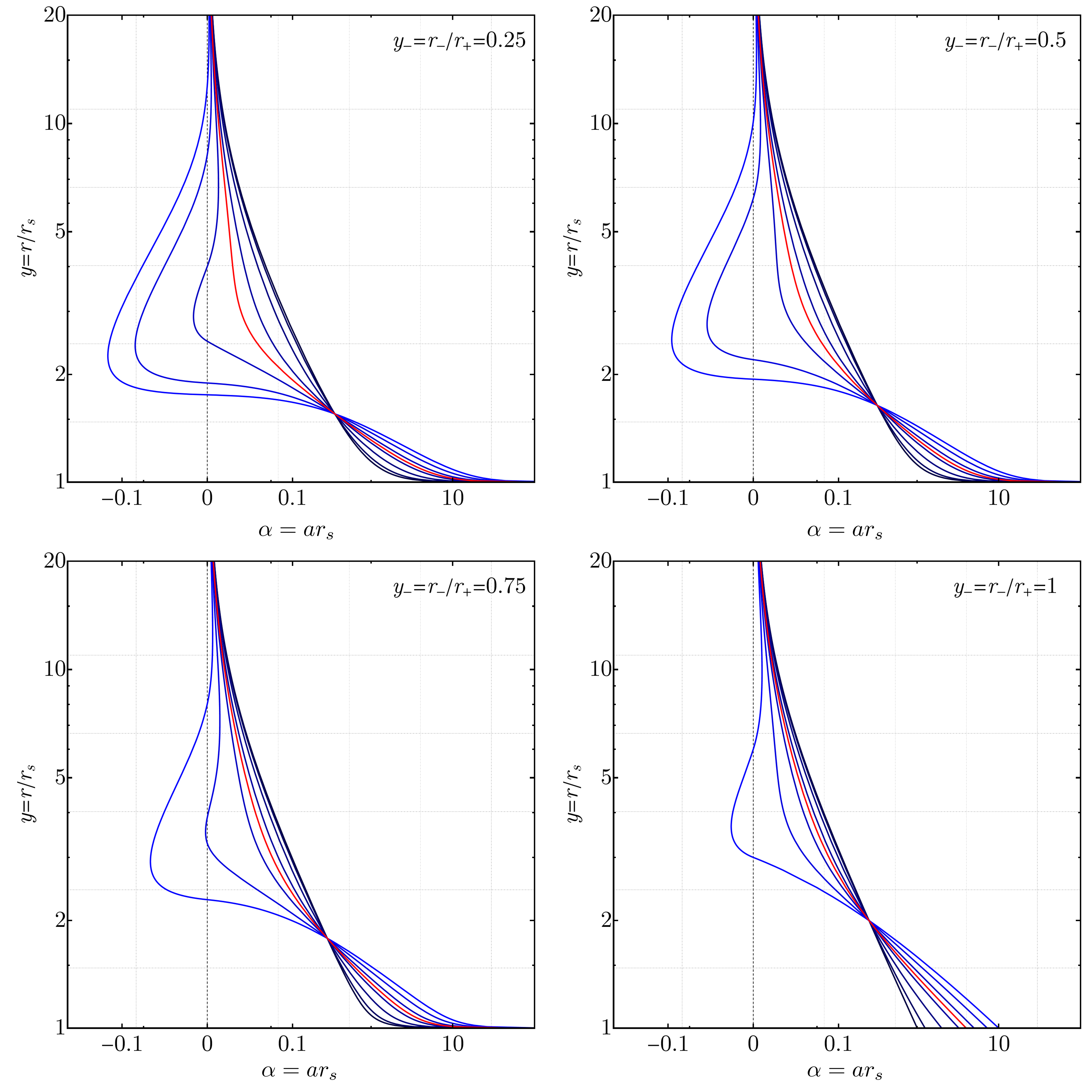} \caption{Dependence of the
position of the circular orbit on the external force for various $y_{-}%
=r_{-}/r_{+}$, where $r_{\mp}$ are the inner and outer horizons, respectively
and for various values of angular momentum $b$. Color of the curve corresponds
to the same value of $b$ as on FIG. 1.}%
\label{fig_5}%
\end{figure}

Let us start with the influence of charge on the position of circular orbits,
depending on the force (see FIG. \ref{fig_5}) 

\begin{figure}[ptb]
\centering
\includegraphics[width=1\linewidth]{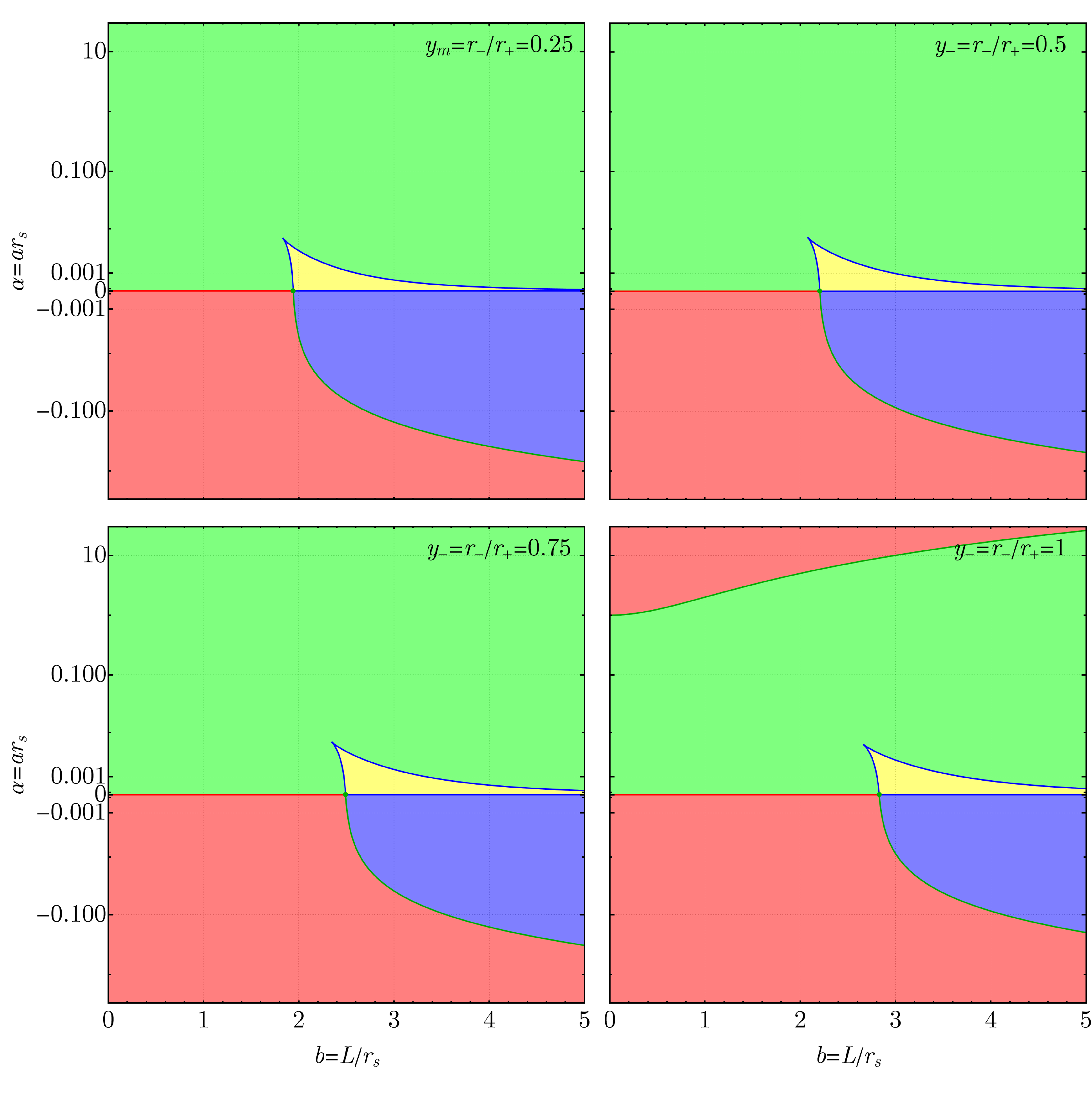} \caption{Diagrams representing
the number of possible circular orbits depending on the force $f=Fr_{+}$ and
angular momentum $b=L/r_{+}$ for various $y_{-}=r_{-}/r_{+}$. Colors of
different represent various amounts of roots: red-no roots, green-1 root,
blue-2 roots, yellow-3 roots.}%
\label{fig_6}%
\end{figure}

On these plots, an interesting behavior can be spotted. Namely, with the
increase of $y_{-}$ at the region away from the horizon ($y\gtrsim1.1$), the
force required to keep a particle at a given radius also increases. Moreover,
this phenomenon is observed for any value of $b$. However, if one looks at the
region close to the horizon, one can notice that the rate of divergence of the
force for $y\rightarrow1$ decreases with the increase of $y_{-}$. The extremal
case $y_{-}=1$ is considered below.

\begin{figure}[ptb]
\centering
\includegraphics[width=1\linewidth]{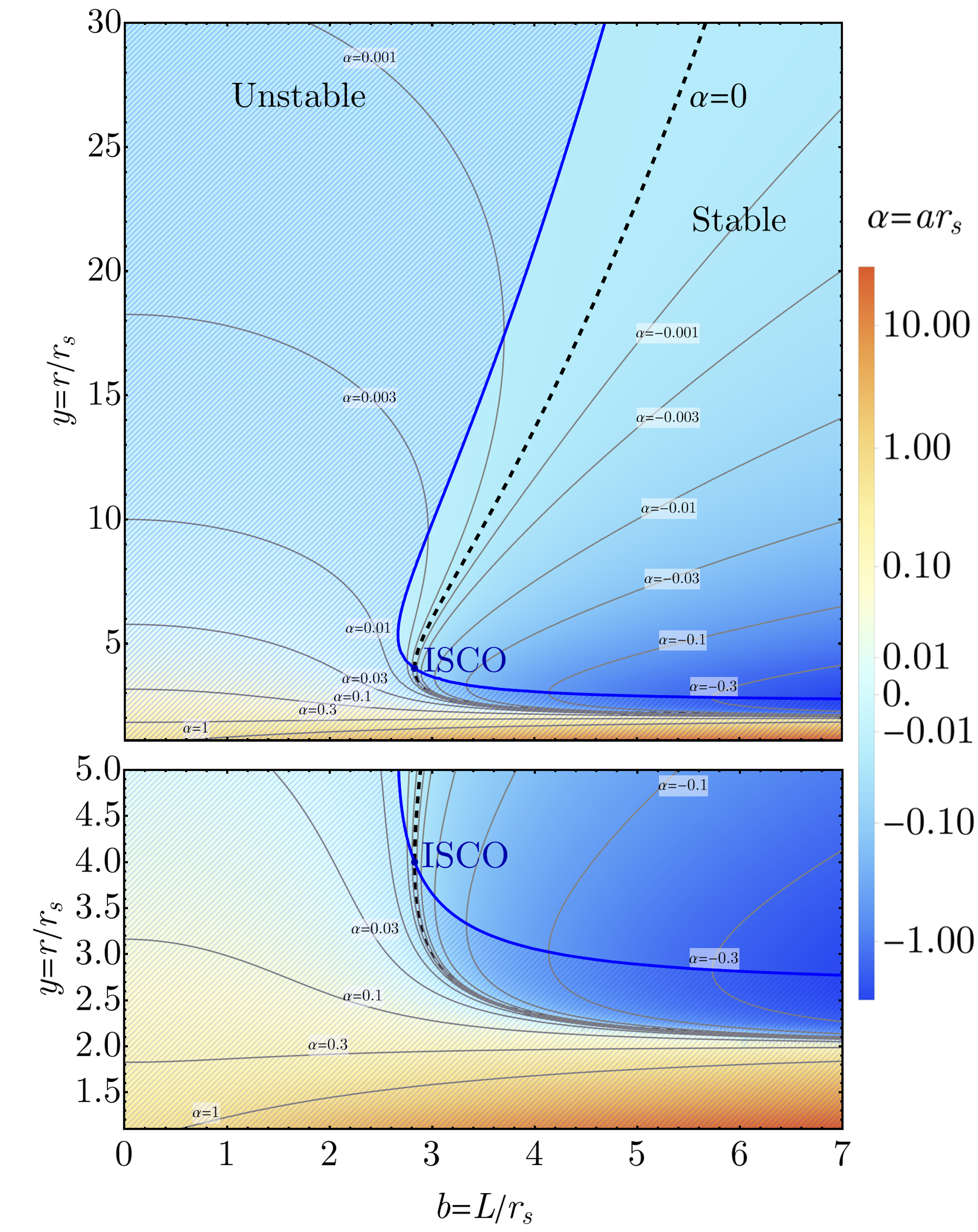} \caption{Stability diagram for
the extremal Reissner-Nordstrom black hole. }%
\label{fig_7}%
\end{figure}

The number of possible circular orbits for the given $y_{-}$ is presented in
FIG. \ref{fig_6}. One can notice that with the increase of $y_{-}$, the value
of $b$, required to have two or three roots, becomes higher. However,
qualitatively, the diagram remains the same. The only significant difference
happens for the extremal case, where an upper limit for the force (that was
not present for a non-extremal black hole) appears. This is in agreement with
the statements of Sec. \ref{hor} where we pointed out that for the extremal
black hole the horizon value of acceleration is finite.

Also, we would like to investigate is the stability diagram (FIG.
\ref{fig_7}). Here we present only the extreme case because the non-extremal
RN solution only slightly differs from the Schwarzschild solution, and the
only qualitatively interesting things appear only for the extremal case. From
FIG. \ref{fig_7}, one can clearly see that the range of angular momenta and
radial positions where the circular orbit is stable becomes broader. Moreover,
the ISCO trajectory can be obtained for bigger values of both $y$ and $b$.
What is interesting is that any circular particle close to the horizon remains
unstable even though the force required to keep such a particle remains finite.

\begin{figure}[ptb]
\centering
\includegraphics[width=1\linewidth]{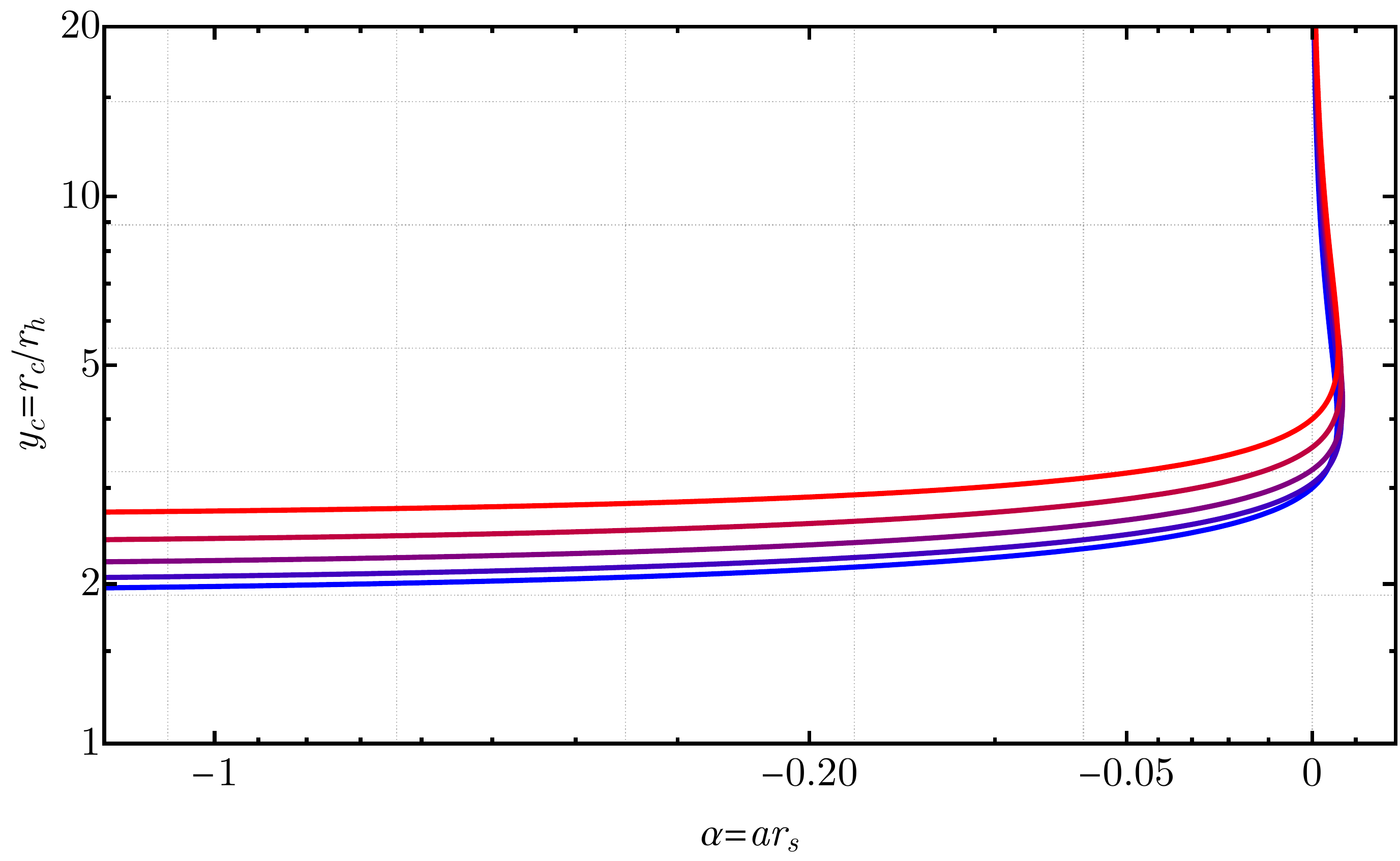}
\caption{Position of ISCO depending on the force acting on the particle in
Reissner-Nordstr\"{o}m spacetime. Here, different colors correspond to
different parameters $y_{-}$. From blue (bottom) to red (top): $y_{-}=0,~y_{-}=0.25,~y_{-}%
=0.5,~y_{-}=0.75,~y_{-}=1$.}%
\label{pic2_isco}%
\end{figure}

The last thing we wish to investigate are the ISCO particles in the
Reissner-Nordstr\"{o}m background. The resulting dependencies of the position
of the ISCO on the force are presented in Fig. \ref{pic2_isco}. On this plot
there are 5 curves that correspond to different parameters $y_{-}$, namely
from blue curve to the red $y_{-}$ takes values $y_{-}=0,~y_{-}=0.25,~y_{-}%
=0.5,~y_{-}=0.75,~y_{-}=1$. These curves were obtained by the same algorithm
as described in Sec. \ref{schw}.

We can notice that as $y_{-}$ increases (which means that the black hole is
closer to extremality), the corresponding curves have the same properties as
the one in Fig. \ref{pic1_isco}, namely that for negative forces there exists
a minimal value of radius for the ISCO particle, for a small positive force
there are two ISCO trajectories and there exists a maximal limit of the force
above which ISCO trajectories cannot exist. What is interesting is that as a
black hole reaches extremality, the minimal ISCO radius increases and reaches
$y=8/3$ for the extremal BH. Also, the position where the maximal force is
achievable is shifted upward, reaching $y=16/3$ for extremal BH. Thus, we see
that basically the presence of the new parameter (namely, the charge of a
black hole) does not change the dependence of the position of ISCO
qualitatively, only some numerical values change.

\subsection{RN metric: extremal horizon and fake trajectories}

\label{rnext}

If $r_{+}=r_{-}$, so $y_{-}=1$, the horizon becomes extremal. In this case,
eq. (\ref{f_y_rn}) turns into%
\begin{equation}
\alpha=ar_{+}=\frac{y^{2}-b^{2}y+2b^{2}}{y^{4}}\text{.} \label{fextr}%
\end{equation}

Formally, there is a solution of this equation corresponding to the horizon
($y=1$), provided $\alpha=1+b^{2}$.

However, the trajectory $r=r_{+}=const$ is fake since a massive particle has a
timelike trajectory that cannot lie on the horizon which is a lighltike
surface. Its fake appearance is due to the use of coordinates $t$, $r$ that
are inadequate on the horizon. To repair this drawback, let us pass, say, to
the Gullstrand-Painlev\'{e} coordinates \cite{pain}, \cite{gull} according to%
\begin{equation}
dt=d\tilde{t}-\frac{\sqrt{1-A}dr}{A}\text{.}%
\end{equation}
Then,%
\begin{equation}
ds^{2}=-Ad\tilde{t}^{2}+dr^{2}+2d\tilde{t}dr\sqrt{1-A}+r^{2}d\omega
^{2}\text{.}%
\end{equation}

In the coordinates $\tilde{t},r$ the metric is regular on the horizon. Along
the trajectory $r=r_{+}$,%
\begin{equation}
ds^{2}=r_{+}^{2}d\phi^{2}\text{.}%
\end{equation}

It is spacelike, if $L\neq0$, or lightlike, if $L=0$. But it cannot be
timelike. The situation is similar to that for the Kerr metric \cite{circkerr}
(Sec. III C) or generic stationary axially symmetric black hole spacetimes
\cite{zas15} (Section VI).

What happens, if $\alpha=1+b^{2}$ exactly? It follows from eq. (\ref{fextr})
that%
\begin{equation}
x[x^{3}(1+b^{2})+4x^{2}(1+b^{2})+x(6b^{2}+5)+2+5b^{2}]=0\text{,}%
\end{equation}
where $x=y-1$.

Obviously, this equation does not have roots outside the horizon, so circle
orbits are absent.

Let $\alpha=1+b^{2}-\delta$, where $0<\delta\ll1$. Then,%
\begin{equation}
x\approx\frac{\delta}{2+5b^{2}}\text{.} \label{xd}%
\end{equation}

Thus circle orbits do exist in the vicinity of the horizon in this case. The
acceleration remains finite in accordance with Sec. \ref{hor}.

\subsection{RN metric: near--extremal horizons}

\label{rnne}

A separate questions arises about the existence of the near-horizon ISCO. The
difference between pure extremal and almost extremal black holes may be
essential since, say, for the Kerr metric, such orbits do exist in the second
case \cite{72}. However, in our context, it follows from (\ref{bex}) that for
a pure extremal horizon, $b^{2}$ for the ISCO would become negative. If a
horizon is near-extremal, small corrections cannot change this situation.
Therefore, we conclude that the constant force cannot play a role similar to
rotation, and the near-horizon ISCO is absent.

However, there is another question: can circle trajectories (not ISCO) exist
near the horizon with finite acceleration? For the nonextremal metric they are
absent, for extremal horizons they are possible. What happens in the
intermediate situation when a black hole is not extremal but is very close to
the extremal state? To elucidate this issue, let us allow acceleration to vary
and instead of a constant acceleration, in this subsection we consider
$\alpha=\alpha(y)$. In our treatment above, we fixed $\alpha$ and looked for
$y$ that corresponds to this value of $\alpha$. Now, we invert the situation:
we fixed $y$ and scrutiny the properties of corresponding $\alpha$, i.e. the
force necessary to keep a circular particle on this radius.

We are interested in the near-extremal horizon, so we put%
\begin{equation}
y_{-}=1-\varepsilon\text{, }0<\varepsilon\ll1\text{.}%
\end{equation}
In the near-horizon region
\begin{equation}
y=1+\delta\text{, }0<\delta\ll1.
\end{equation}

Then, it follows from (\ref{f_y_rn}) that%
\begin{equation}
\alpha\approx\frac{(1+b^{2})(2\delta+\varepsilon)}{2\sqrt{\delta}%
\sqrt{\varepsilon+\delta}}\text{.}%
\end{equation}
Let us consider the relation $\delta=h\varepsilon$, where $h>0$ is some
number. Then, acceleration on the near-horizon circle orbit%
\begin{equation}
\alpha\approx\frac{(1+b^{2})(2h+1)}{2\sqrt{h(h+1)}}\text{.} \label{h}%
\end{equation}
For any finite $h$ the quantity $\alpha$ is finite. Thus near-horizon circle
orbits with finite acceleration do exist. In the limit $h\rightarrow\infty$
that corresponds to $\varepsilon\ll\delta$ we return, neglecting terms of the
order $\delta$, to the extremal case considered above with $\alpha=1+b^{2}$ on
the horizon (but with the reservation that a corresponding circle trajectory
on the horizon is fake).

\section{High energy collisions\label{high}}

In addition to the analysis we did so far for a single circular particle,
there is an astrophysical interest in investigating collisional phenomena with
circular particles. To this end, let us assume that two particles, $1$ and $2$
collide in point $C$. The corresponding collision energy in the center of mass
frame is given by%
\begin{equation}
E_{c.m.}^{2}=-(m_{1}u_{\mu}^{(1)}+m_{2}u_{\mu}^{(2)})(m_{1}u^{(1)\mu}%
+m_{2}u^{(2)\mu})=m_{1}^{2}+m_{2}^{2}+2m_{1}m_{2}\gamma,\text{ \ \ }%
\gamma=-u_{\mu}^{(1)}u^{\mu(2)}\text{.}%
\end{equation}

It follows from equations of motion (\ref{ut}) that%
\begin{equation}
\gamma=\frac{X^{(1)}X^{(2)}-\sigma P^{(1)}P^{(2)}}{N^{2}}-\frac{\mathcal{L}%
^{(1)}\mathcal{L}^{(2)}}{r^{2}} \label{ga}%
\end{equation}
where $\sigma=+1$ if particles move in the same direction and $\sigma=-1$ if
they do it in opposite ones, $r_{+}$ is the point of collision.

In the intermediate case of circle particle 1 $P^{(1)}=0$, so%
\begin{equation}
\gamma=\frac{X^{(1)}X^{(2)}}{N^{2}}\text{.}%
\end{equation}

Let the RN metric be a pure extremal. According to (\ref{xd}), in this case
there exist circular orbit close to the horizon, with small $\delta$, where
$f=1+b^{2}-\delta$ and $y=\frac{r}{r_{+}}=1+x$, $X^{(1)}=X(r_{c}%
)=mN(r_{c})\sqrt{1+\dfrac{\mathcal{L}_{1}^{2}}{r_{c}^{2}}}$ according to
(\ref{z})$.$ Then, assuming that particle 2 is usual (not fine-tuned, so
$X^{(2)}=O(1)$ on the horizon) $\gamma\sim N_{c}^{-1}\sim\delta^{-1}$. A
similar result is valid for near-horizon circular particles with
$\varepsilon\sim\delta$ in the near-extremal case considered above. Thus high
energy collisions are possible near extremal and near-extremal RN black holes.

\newpage

\section{Summary and conclusions\label{sum}}

In this work, we developed a general approach to properties of circular orbits
of massive particles in spherically symmetric static spacetimes under the
action of the external force. Our analysis can be applied to any physically
important setup, as we do not specify the nature of the external force and
study how the \textit{arbitrary} force affects circular orbits. In particular,
we considered the Schwarzschild and Reissner-Nordstr\"{o}m (RN) metrics. In
doing so, we defined the ISCO with a force taken into account and, for a
constant force, found the regions of stability. We also traced, how the change
of a force leads to the appearance of new branches of circle orbits.

We showed that near-horizon circle trajectories exist for the extremal and
near-extremal RN metric. In doing so, the force required to keep a particle,
remains finite in the horizon limit although it diverges for the nonextremal
metric. However, it is demonstrated that such orbits are not stable, and there
is no way of having ISCO particles close to the horizon. Also, we show that
for pure extremal horizons, the circle orbits do not exist for free particles
but such trajectories become possible if a force is taken into account.

We considered high-energy particle collisions for the near-extremal and
extremal RN metric, where it was shown that for the circular particles
orbiting close to the horizon (their existence was proven in Sections
\ref{rnext} and \ref{rnne}), the collision energy may significantly increase
and is proportional to $\delta^{-1}$, where $\delta$ represents how close the
particle is to horizon.

We hope that our results may be useful for the investigations of various
astrophysical processes with circular particles near the horizons of black holes.

\section{Data Availability Statement}
This manuscript has associated data
in a data repository [https://zenodo.org/records/19252481].

\appendix

\section{Calculation of acceleration in the comoving frame}

\label{append_1}

In this section, we obtain the acceleration of a particle moving along the
circular orbit, in the comoving frame. For brevity, we will call it ``circular particle".

Let such a particle move in the equatorial plane of the spherically symmetric
spacetime (\ref{metr}). For it, $r=const,$ $\theta=const,$ and during the
motion only $t$ and $\varphi$ coordinates change.

\subsection{Frames and tetrads}

\qquad Coordinate components of the 4-acceleration are given by (\ref{accel}).

For our computations we also require tetrad components of acceleration
$a^{(a)}=e_{\mu}^{(a)}a^{\mu}$ in two frames. One of them is the static one to
which we attach the tetrad
\begin{align}
e_{(0)}^{\mu}  &  =\dfrac{1}{N}(1,0,0,0),~~~e_{(1)}^{\mu}=\sqrt{A}%
(0,1,0,0),\nonumber\\
e_{(2)}^{\mu}  &  =\dfrac{1}{r}(0,0,1,0),~~~e_{(3)}^{\mu}=\dfrac{1}%
{r\sin\theta}(0,0,0,1) \label{ozamo_fr}%
\end{align}

In this tetrad frame, the four-velocity reads (for circular particles $X$ is not independent and at a given $r_c$ is given by (\ref{z}))
\begin{equation}
u^{(a)}=\left(  \frac{X}{mN},0,0,\frac{\mathcal{L}}{r}\right)  \text{.}
\label{utet}%
\end{equation}

According to (\ref{ats}), (\ref{ar1}), the components of acceleration of a
circle particle in the static frame are given by%
\begin{align}
a_{s}^{(0)}  &  =0,\label{ao_0}\\
a_{s}^{(1)}  &  =\dfrac{X}{m^{2}}\frac{\sqrt{A}}{N^{2}}\partial_{r}%
X,\label{ao_1}\\
a_{s}^{(2)}  &  =0,\label{ao_2}\\
a_{s}^{(3)}  &  =0 \label{ao_3}%
\end{align}

We use letter \textquotedblleft s" to indicate that such an observer is
static. During this computation we took into account that the system has a
symmetry with respect to the equatorial plane, and derivatives $\partial
_{\theta}$ of all metric functions are thus zero.

\subsection{Relationships between different frames}

In general, the static (S) frame becomes singular near the horizon whereas a
more natural one consists in attaching the tetrad to a comoving frame (CF). To
establish transformations between the static and CF frames, we have to find
the three-velocity in the static frame. Direct computation gives us%
\begin{equation}
V^{(i)}=-\frac{e_{\mu}^{(i)}u^{\mu}}{e_{\mu}^{(0)}u^{\mu}}=\left(  0,\frac
{LN}{rX}\right)  . \label{V}%
\end{equation}

The absolute value of the three-velocity is
\begin{equation}
|V|=\frac{LN}{r|X|}=\frac{L}{\sqrt{r^{2}+L^{2}}} \label{VNX}%
\end{equation}
where for definiteness we assumed $L>0$ (in the last equality we used
(\ref{z})). To transform to comoving frame, one has to perform a boost%
\begin{align}
\widetilde{e}_{(0)}  &  =\gamma(e_{(0)}+|V|e_{(3)}),\text{ \ \ }\widetilde
{e}_{(1)}=e_{(1)}^{\prime\prime},\label{b0}\\
\widetilde{e}_{(3)}  &  =\gamma(e_{(3)}+|V|e_{(0)}),\text{ \ \ }\widetilde
{e}_{(2)}=e_{(2)}^{\prime\prime}, \label{b1}%
\end{align}

where $\gamma=\frac{1}{\sqrt{1-V^{2}}}=\sqrt{1+\frac{L^{2}}{r^{2}}}.$

Combining all these transformations, one obtains that comoving tetrad is given
by%
\begin{equation}
\left(
\begin{array}
[c]{c}%
\widetilde{e}_{(0)}\\
\widetilde{e}_{(1)}\\
\widetilde{e}_{(2)}\\
\widetilde{e}_{(3)}%
\end{array}
\right)  =\left(
\begin{array}
[c]{cccc}%
\sqrt{1+\frac{L^{2}}{r^{2}}} & 0 & 0 & \frac{L}{r}\\
0 & 1 & 0 & 0\\
0 & 0 & 1 & 0\\
\frac{L}{r} & 0 & 0 & \sqrt{1+\frac{L^{2}}{r^{2}}}%
\end{array}
\right)  \left(
\begin{array}
[c]{c}%
e_{(0)}\\
e_{(1)}\\
e_{(2)}\\
e_{(3)}%
\end{array}
\right)  \label{table}%
\end{equation}

or, explicitly,%
\begin{align}
\widetilde{e}_{\mu}^{(0)}  &  =\left(  \sqrt{1+\frac{L^{2}}{r^{2}}%
},0,0,-L\right)  ,\label{e_0_tild}\\
\widetilde{e}_{\mu}^{(1)}  &  =\left(  0,\frac{1}{\sqrt{A}},0,0\right)  ,\\
\widetilde{e}_{\mu}^{(2)}  &  =\left(  0,0,r,0\right)  ,\\
\widetilde{e}_{\mu}^{(3)}  &  =\left(  -\frac{LN}{r},0,0,\sqrt{r^{2}+L^{2}%
}\right)  . \label{e_3_tild}%
\end{align}

One can check that in this frame holds%
\begin{equation}
\widetilde{V}^{i}=-\frac{\widetilde{e}_{\mu}^{(i)}u^{\mu}}{\widetilde{e}_{\mu
}^{(0)}u^{\mu}}=(0,0,0)\text{,}%
\end{equation}
as it should be, and the four-velocity%
\begin{equation}
\widetilde{u}^{(a)}=\widetilde{e}_{\mu}^{(a)}u^{\mu}=(1,0,0,0).
\end{equation}

The components of acceleration in the comoving frame $a_{C}^{(a)}%
=\widetilde{e}_{\mu}^{(a)}a_{s}^{\mu}$ are equal to%
\begin{align}
a_{C}^{(0)}  &  =0,\label{af_0}\\
a_{C}^{(1)}  &  =\dfrac{X}{m^{2}}\frac{\sqrt{A}}{N^{2}}\partial_{r}%
X,\label{af_1}\\
a_{C}^{(2)}  &  =0,\label{af_2}\\
a_{C}^{(3)}  &  =0. \label{af_3}%
\end{align}

As one can see, in both the S and CF frames the only non-zero component of the
acceleration is the radial one. Also, we wish to notice that by using
(\ref{z}), one obtains exactly (\ref{a}) for the acceleration of the
circular particle.

\section{Supplementary calculations for ISCO}

\label{app}

In this Appendix, we give a derivation of some details concerning the behavior
of ISCO for the Schwarzschild and RN metrics

\subsection{Schwarzschild metric}

For the ISCO, $\dfrac{\partial\alpha}{\partial y}=0$. Then, it follows from
(\ref{f_prime}) that on ISCO%
\begin{equation}
b^{2}=\frac{y^{2}(4y-3)}{12y^{2}-34y+21} \label{bi}%
\end{equation}
and%
\begin{equation}
\alpha=\frac{2(y-3)\sqrt{y-1}}{y^{3/2}(12y^{2}-34y+21)}. \label{aly}%
\end{equation}

When $y<3$, $\alpha<0$. The value $y=3$ corresponds just to the case of a free
particle, $\alpha=0$. If $y>3$, two different $y$ correspond to a given
$\alpha$. In particular, if $\alpha\rightarrow+0$, eq. (\ref{small}) for the
second root is valid. The region with two roots $y(\alpha)$ exists for
$0<\alpha<\alpha_{\max}=0.00562435$.

The denominator in (\ref{bi}) vanishes for%
\begin{equation}
y_{1}=\frac{17}{12}+\frac{\sqrt{37}}{12}%
\end{equation}
and%
\begin{equation}
y_{2}=\frac{17}{12}-\frac{\sqrt{37}}{12}.
\end{equation}

As $y_{2}<1$, it lies beyond the horizon and is of no interest for us. When
$y\rightarrow y_{1}+0$, $b^{2}\rightarrow\infty$ and according to
(\ref{f_y_eq}). As $y_{1}>\frac{3}{2}$, the quantity $\alpha$ is negative and
$\alpha\rightarrow-\infty$.

\subsection{Reissner-Nordst\"{o}m metric}

General expression (\ref{f_y_rn}) for $\alpha$ in the case of the RN metric is
rather cumbersome. To avoid long formulas, here we restrict ourselves to the
extremal metric when eq. (\ref{fextr}) is valid. For readers' convenience, we
repeat it below%
\begin{equation}
\alpha=\frac{y^{2}-b^{2}y+2b^{2}}{y^{4}}\text{.} \label{ext}%
\end{equation}

Then,%
\begin{equation}
\dfrac{\partial\alpha}{\partial y}=\frac{-2y^{2}+3b^{2}y-8b^{2}}{y^{5}}.
\label{a1}%
\end{equation}
Following the same line as for the Schwarzschild metric, we obtain that on the
ISCO%
\begin{equation}
b^{2}=\frac{2y^{2}}{3y-8}\text{,} \label{bex}%
\end{equation}%
\begin{equation}
\alpha=\frac{(y-4)y^{2}}{y^{4}(3y-8)}.
\end{equation}
When
\[
y\rightarrow y_{1}+0,~~~y_{1}=\dfrac{8}{3}%
\]
$b\rightarrow\infty$ and $\alpha\rightarrow-\infty$.

The total derivative on ISCO due to $\frac{\partial\alpha}{\partial y}=0$ is
equal to%
\begin{equation}
\frac{d\alpha}{dy}=\dfrac{\partial\alpha}{\partial b^{2}}\frac{db^{2}}%
{dy}=\frac{2(2-y)(3y-16)}{y^{4}(3y-8)^{2}}\text{.}%
\end{equation}

Then, for $y\rightarrow\dfrac{16}{3}$ we see that indeed $\dfrac{d\alpha}%
{dy}=0$ in accordance with the number given above in the main text. In doing
so, $b^{2}\rightarrow\dfrac{64}{9}$, $\alpha\rightarrow\alpha_{\max}=\dfrac
{3}{512}$. The double root $y(\alpha)$ exists for \thinspace$0<\alpha
<\alpha_{\max}$.

\end{document}